\documentclass[aps,prd,amsfonts,amssymb, amsmath, showpacs, showkeys, a4paper, nofootinbib, superscriptaddress, 11pt, raggedbottom]{revtex4-2}

\usepackage{braket}
\usepackage{bm}
\usepackage{graphicx}
\usepackage{slashed}
\usepackage{subfig}
\usepackage{mathrsfs}
\usepackage{color}
\usepackage{mathtools}
\usepackage[normalem]{ulem}
\usepackage{simplewick}

\begin{document}

\title{\Large Frequency shifts induced by light scalar fields}

\author{Christian K\"{a}ding}
\email{christian.kaeding@tuwien.ac.at}
\affiliation{Atominstitut, Technische Universit\"at Wien, Stadionallee 2, 1020 Vienna, Austria}

\begin{abstract}
Light scalar fields are frequently used in modern physics, for example, as candidates for dark energy or dark matter. Open quantum dynamical effects, like frequency shifts, induced by such fields in probe particles used in interferometry experiments might open up new perspectives for constraining such models. In this article, we consider a probe scalar particle as a rough approximation for an atom in matter wave interferometry and discuss the frequency shifts induced by interactions with an environment comprising either one of two screened scalar field models: chameleons or symmetrons. For the $n=-4$ chameleon, we revise a previously obtained expression for the induced frequency shift, but confirm that it can likely not be used to obtain new constraints. However, for symmetrons, we find that induced frequency shifts have the potential to
tightly constrain previously unreachable parts of the parameter space.
\end{abstract}

\keywords{light scalar fields, chameleons, symmetrons, frequency shifts}

\maketitle



\section{Introduction}

In order to address striking issues in modern cosmology, for example, the natures of dark matter and dark energy, modifications of general relativity like scalar-tensor theories have been devised  \cite{Fujii2003,Clifton2011,Joyce2014}. Many of these theories introduce additional scalar fields, which often couple universally to other matter. The resulting gravity-like fifth forces that we would expect to see from such couplings have already been tightly constrained within our Solar System \cite{Dickey1994,Adelberger2003,Kapner2007}. However, scalar field models with screening mechanisms (screened scalar fields), for example, the well-known chameleons \cite{Khoury2003,Khoury20032} and symmetrons \cite{Dehnen1992, Gessner1992, Damour1994, Pietroni2005, Olive2008, Brax2010,Hinterbichler2010,Hinterbichler2011}, can circumvent such constraints by rendering their fifth forces feeble in environments at least as dense as our Solar System. Consequently, they are of particular phenomenological interest and actively searched for in a variety of experiments; see Refs.~\cite{Fischer:2024eic,Burrage:2017qrf} for current constraints. Examples for particularly interesting aspects of these models that motivate the search for them are the facts that chameleons naturally appear in reinterpretations of $f(R)$-gravity as scalar-tensor theories \cite{Faulkner:2006ub}, and symmetron fifth forces have recently been discussed as promising alternatives to particle dark matter \cite{Burrage:2016yjm,OHare:2018ayv,Burrage:2018zuj,Kading:2023hdb}.

Matter wave interferometry is one of the most successful ways of constraining screened scalar fields. However, as of yet, interferometric experiments have been treating the scalar fields only as classical backgrounds; see, e.g., Refs.~\cite{Burrage:2014oza,Hamilton:2015zga,Lemmel:2015kwa,Burrage:2015lya,Elder:2016yxm,Burrage:2016rkv,Jaffe:2016fsh,Sabulsky:2018jma,Fischer:2023eww,Mueller2024}. For this reason, such experiments could not harness the potential of open quantum dynamical effects, like decoherence or frequency shifts, induced by interactions between the matter waves and an environment comprising screened scalar fields. Consequently, Refs.~\cite{Burrage2018,Kading:2023mdk} have initiated a theoretical discussion of how light scalar fields could influence the quantum dynamics of cold atoms in interferometry experiments. However, while Ref.~\cite{Burrage2018} has a strong focus on developing the necessary mathematical framework, it discusses only one exemplary point in the $n=-4$ chameleon parameter space, and Ref.~\cite{Kading:2023mdk} deals solely with the environment-dependent dilaton model \cite{Damour1994,Gasperini:2001pc,Damour:2002nv,Damour:2002mi,Brax:2010gi,Brax:2011ja,Brax2022,Fischer:2023koa}. In both articles, frequency shifts turned out to be the dominant induced open quantum dynamical effects.

In this article, we complement the investigations of Refs.~\cite{Burrage2018,Kading:2023mdk} by extending and correcting the discussion about frequency shifts induced by $n=-4$ chameleons, and by investigating symmetrons for the very first time in this context. At first, we recapitulate the definitions and essential properties of these two models. Subsequently, we employ the methods developed in Ref.~\cite{Kading2022x} in order to predict frequency shifts induced by those models. For the $n=-4$ chameleon model, we obtain a new expression for the frequency shift that differs from the one in Ref.~\cite{Burrage2018}, and explain that the old result was not correct because its time dependence was not properly taken into account when deriving it. We then find that it could be possible to constrain this chameleon model by looking at its induced frequency shifts, but such constraints would only be on parts of the parameter space that have already been covered by other experiments. Finally, we look at the symmetron model that has, as of yet, not been discussed in this context. Interestingly, we find that symmetron-induced frequency shifts have the potential to actually constrain regions of the model's parameter space that have never been reached before. 


\section{Chameleons and symmetrons}
\label{sec:Models}

In this article, we focus on two particular screened scalar field models: $n=-4$ chameleons and symmetrons. In the Einstein frame \cite{Fujii2003}, both models can be described by the action \cite{Fischer:2024gni}
\begin{eqnarray}\label{eq:Action}
S &=& \int  d^4x\, \sqrt{-g} \left( \frac{m_\text{Pl}^2}{2}\,R - \frac{1}{2}(\partial\varphi_\alpha)^2 - V_\alpha(\varphi_\alpha) \right) + \int d^4x\,\sqrt{-\tilde g^\alpha}\,\mathcal{L}_\text{SM} (\tilde g_{\mu \nu}^\alpha,\psi)~,
\end{eqnarray}
where $m_\text{Pl}$ is the reduced Planck mass; $\varphi_\alpha$ denotes the screened scalar field; $\alpha \in \{C,S \}$ labels the particular model, i.e., chameleon ($C$) or symmetron ($S$); $V_\alpha(\varphi_\alpha)$ is the self-interaction potential of model $\alpha$; and $\mathcal{L}_\text{SM}$ represents the Lagrangian for the Standard Model field $\psi$. For brevity's sake, the Jordan frame metric $\tilde g^\alpha_{\mu \nu} = A_\alpha^2(\varphi_\alpha)g_{\mu \nu}$ with conformal factor $A_\alpha(\varphi_\alpha)$ was kept in Eq.~(\ref{eq:Action}). Note that Eq.~(\ref{eq:Action}) can also describe other scalar-tensor theories with light scalar fields, including the environment-dependent dilaton \cite{Damour1994,Gasperini:2001pc,Damour:2002nv,Damour:2002mi,Brax:2010gi,Brax:2011ja,Brax2022,Fischer:2023koa}. The universal coupling to the matter sector induces a fifth force that must be screened in environments at least as dense as our Solar System in order to circumvent current constraints. Chameleons and symmetrons each have their own way of realising such a screening. However, in both cases, Eq.~(\ref{eq:Action}) gives rise to effective potentials that are key to the fifth force screening:
\begin{eqnarray}\label{eq:EffPot}
    V_{\alpha;\text{eff}} (\varphi_\alpha) &:=& V_\alpha(\varphi_\alpha)- A_\alpha(\varphi_\alpha)T^\nu_{\phantom{\nu}\nu}~,
\end{eqnarray}
where $T^\nu_{\phantom{\nu}\nu}$ is the trace of the stress-energy tensor of $\psi$.

While there are infinitely many possible self-interaction potentials $V_C \sim \varphi_C^{-n}$ with $n \in \mathbb{Z}^+ \cup 2\mathbb{Z}^-\setminus\{-2\}$ for chameleons, we follow Ref.~\cite{Burrage2018} and consider the $n=-4$ chameleon since it is one of the two most popular models (the other one is $n=1$)\footnote{ The cases $n=1$ and $n=-4$ are the most popular models since they have the smallest possible values of $|n|$ for the positive and negative $n$ chameleon models, respectively.} and does not give rise to an inverse potential, which would otherwise cause intricacies in the quantum field theoretical treatment. It is specified by; see, e.g., Ref.~\cite{Fischer:2024gni};
\begin{eqnarray}
\label{eq:ChamA}
    V_C &=& \frac{\lambda_C}{4!}\varphi_C^4~,~A_C(\varphi_C) \,=\, \exp\left(\frac{\varphi_C}{\mathcal{M}_C}\right)\,\approx\, 1 + \frac{\varphi_C}{\mathcal{M}_C} + \frac{1}{2}\frac{\varphi_C^2}{\mathcal{M}_C^2}~,
\end{eqnarray}
where $\lambda_C$ is a dimensionless self-coupling constant; the mass scale $\mathcal{M}_C$ determines the coupling strength to matter; and $\varphi_C \ll \mathcal{M}_C$ is usually assumed. Its effective potential, cf.~Eq.~(\ref{eq:EffPot}), leads to a non-vanishing vacuum expectation value (VEV) and a $T^\nu_{\phantom{\nu}\nu}$-dependent effective mass:
\begin{eqnarray}
    \langle\varphi_C\rangle &=& \left( \frac{6}{\lambda_C} \frac{T^\nu_{\phantom{\nu}\nu}}{\mathcal{M}_C} \right)^{1/3}~,~m_C^2 \,=\, \left[ \frac{9}{2}\lambda_C \bigg(\frac{T^\nu_{\phantom{\nu}\nu}}{\mathcal{M}_C}\bigg)^2 \right]^{1/3}~.
\end{eqnarray}
Clearly, a chameleon field is the more massive the larger the magnitude of $T^\nu_{\phantom{\nu}\nu}$. In non-relativistic environments, $T^\nu_{\phantom{\nu}\nu} \approx - \sigma$  with the density $\sigma$. Consequently, a chameleon fifth force is short-ranged, i.e.,~screened, in sufficiently dense matter. 

Symmetrons are defined by; see, e.g., Ref.~\cite{Fischer:2024gni};  
\begin{eqnarray}
\label{eq:SymA}
    V_S &=& -\frac{\mu^2}{2}\,\varphi_S^2 + \frac{\lambda_S}{4}\,\varphi_S^4~,~A_S(\varphi_S) \,=\, 1 + \frac{\varphi_S^2}{2\mathcal{M}_S^2}~,
\end{eqnarray}
where $\mu$ is a tachyonic mass; $\lambda_S$ is a dimensionless constant; and $\mathcal{M}_S$ is the mass scale that quantifies the coupling to matter. From the resulting effective potential, cf.~Eq.~(\ref{eq:EffPot}), we can derive 
\begin{eqnarray}
    \langle\varphi_S\rangle &=&
    \begin{cases}
    0 &,~ -T^\nu_{\phantom{\nu}\nu} \geq \mu^2 \mathcal{M}_S^2
    \\
   \pm \sqrt{\frac{1}{\lambda_S}\left( \mu^2 + \frac{T^\nu_{\phantom{\nu}\nu}}{\mathcal{M}_S^2} \right)}&,~ -T^\nu_{\phantom{\nu}\nu} < \mu^2 \mathcal{M}_S^2
    \end{cases}
~,~~~~
    m_S^2 \,=\, 
    \begin{cases}
    -\frac{T^\nu_{\phantom{\nu}\nu}}{\mathcal{M}_S^2}-\mu^2 &,~ -T^\nu_{\phantom{\nu}\nu} \geq \mu^2 \mathcal{M}_S^2
    \\
    2\left(\mu^2+\frac{T^\nu_{\phantom{\nu}\nu}}{\mathcal{M}_S^2} \right)&,~ -T^\nu_{\phantom{\nu}\nu} < \mu^2 \mathcal{M}_S^2
    \end{cases}~.~~~~
\end{eqnarray}
At leading order, the symmetron fifth force is proportional to $\langle\varphi_S\rangle$, which implies a fifth force-screening for situations in which $-T^\nu_{\phantom{\nu}\nu} \geq \mu^2 \mathcal{M}_S^2$, i.e.,~in sufficiently dense environments.


\section{Frequency shifts}

We are interested in studying frequency shifts induced in an atom interferometer by interactions with an environment comprising hypothetical chameleon or symmetron fields. The most natural way of describing quantum effects induced by fields is quantum field theory. For this reason, we want to employ the quantum field theoretical method presented in Ref.~\cite{Kading2022x}. However, since complex composite objects like atoms are difficult to discuss in such a framework, we make a rough approximation by considering a real scalar field $\phi$ as a proxy for a cold atom. Such a strong simplification will of course not allow us to make predictions that we can fully compare to real experimental data. However, when considering screened scalar fields as an environment, approximating an atom by a scalar field is not as far-fetched as it might initially seem. This is due to the fact that, for a screened scalar field, the atom is simply a blob of mass mostly concentrated in the atom's nucleus since screened scalar fields only couple to the trace of the atom’s stress-energy tensor. Any other properties of atoms or their nuclei like spin or charge are negligible for screened scalar fields. Consequently, and since a nucleus contains the by far largest part of an atom's mass but is much smaller than the atomic radius, a cold atom or rather its nucleus can be considered as a real scalar particle as a first approximation. In this way, we will be able to at least get an idea of the parts of the screened scalar field parameter spaces that could be constrained by looking at induced frequency shift. The same approximation was successfully employed in Refs.~\cite{Burrage2018,Kading:2023mdk}. In order to make this a more realistic approximation, we consider only parts of the model parameter spaces that leave the screened scalar fields unaffected by the presence of the atom, and restrict our discussion to the single-particle state of the atom, i.e., the atom is neither annihilated nor are additional atoms and anti-atoms being produced. 

We can expand the reduced density matrix that describes the open quantum dynamics of $\phi$ under the influence of another scalar field $\varphi_\alpha$ in a momentum basis restricted to the single-particle subspace of Fock space:
\begin{eqnarray}
    \hat{\rho}_\phi(t) &=& \int\frac{d^3p d^3p'}{(2\pi)^64E^\phi_{\mathbf{p}}E^\phi_{\mathbf{p}'}} \rho(\mathbf{p};\mathbf{p}';t) \ket{\mathbf{p};t}\bra{\mathbf{p}';t}~,
\end{eqnarray}
where
$E^\phi_{\mathbf{p}} = \sqrt{M^2 + \mathbf{p}^2}$ is the atom's energy with atom mass $M$, and
$\rho(\mathbf{p};\mathbf{p}';t)  = \bra{\mathbf{p};t}\hat{\rho}_\phi(t) \ket{\mathbf{p}';t}$ are the density matrix elements at time $t$. Note that $M$ is the mass that we assume to have been determined using experimental setups in which any corrections due to the hypothetical presence of screened scalar fields can be neglected.

The interaction-free atom has the unitary dynamics
\begin{eqnarray}
    \rho_\text{free}(\mathbf{p};\mathbf{p}';t) &=& e^{-\mathrm{i}u(\mathbf{p},\mathbf{p}') t}\rho_\text{free}(\mathbf{p};\mathbf{p}';0)~,
\end{eqnarray}
from time $0$ to time $t$, with the frequency $u(\mathbf{p},\mathbf{p}') := E^\phi_{\mathbf{p}} - E^\phi_{\mathbf{p}'}$. From interactions with an environment comprising a scalar field $\varphi_\alpha$, we expect to find frequency shifts $\Delta u_\alpha(\mathbf{p},\mathbf{p}')$, such that 
\begin{eqnarray}
\label{eq:NonUniEv}
    \rho(\mathbf{p};\mathbf{p}';t) &=& e^{-\mathrm{i}\left[u(\mathbf{p},\mathbf{p}') +\Delta u_\alpha(\mathbf{p},\mathbf{p}')\right]t}\rho_\text{free}(\mathbf{p};\mathbf{p}';0)~.
\end{eqnarray}
Of course there could and likely will be decoherence terms, but we do not consider them in our discussion since they are usually subleading compared to the frequency shift terms; see Refs.~\cite{Burrage2018,Kading:2023mdk}. Note that, on the right-hand side of Eq.~(\ref{eq:NonUniEv}), we have the free density matrix element at time $0$. This results from requiring no initial correlations between system and environment, which is usually necessary when employing methods from the theory of open quantum systems \cite{Breuer2002}. 

Following Ref.~\cite{Burrage2018}, we split a screened scalar field into a background $\langle \varphi_\alpha \rangle \neq 0$ and a small fluctuation $\chi_\alpha$, such that $\varphi_\alpha = \langle \varphi_\alpha \rangle + \chi_\alpha$, and define the following Einstein frame actions 
\begin{eqnarray}
\label{eq:FreeAtomAct}
S_{\phi}[\phi] &:=& \int_x \left[ - \frac{1}{2} g^{\mu \nu} \partial_{\mu} \phi \partial_{\nu} \phi - \frac{1}{2}  A_\alpha^2(\langle \varphi_\alpha \rangle)  M^2 \phi^2 \right]~, 
\\
\label{eq:FreeScSc}
S_{\chi_\alpha}[\chi_\alpha] &:=& \int_x \left[ - \frac{1}{2} g^{\mu \nu} \partial_{\mu} \chi_\alpha \partial_{\nu} \chi_\alpha - \frac{1}{2} m_\alpha^2 \chi_\alpha^2 \right]~, 
\\
\label{eq:SelfIntact}
S_{\text{self}}[\chi_\alpha] &:=& \int_{x \in \Omega_t} \left[ -\frac{\lambda_\alpha}{4!} \Big( \chi_\alpha^4 + 4\langle \varphi_\alpha \rangle \chi_\alpha^3 \Big) \right]~,
\\
\label{eq:interact}
S_{\text{int}}[\phi, \chi_\alpha] &:=& \int_{x \in \Omega_t} \left[ -\frac{1}{2} \beta_\alpha M\chi_\alpha \phi^2 -\frac{1}{4}\gamma_\alpha \chi_\alpha^2 \phi^2 \right]~, 
\end{eqnarray}
where
\begin{eqnarray}
    \int_x &:=& \int d^4x~,
\end{eqnarray}
$\Omega_t := [0,t] \times \mathbb{R}^3$, $\beta_\alpha,\gamma_\alpha \ll 1$ are dimensionless coupling constants, and the conformal factor in the mass term of the otherwise free atom action results from the presence of a conformally coupling scalar field; see Ref.~\cite{Burrage2018} for more details. Due this mass rescaling, we introduce a modified energy $E^{\phi,\alpha}_{\mathbf{p}} = \sqrt{A_\alpha^2(\langle \varphi_\alpha \rangle)M^2 + \mathbf{p}^2}$. Note that Eqs.~(\ref{eq:FreeAtomAct})-(\ref{eq:interact}) are valid for both screened scalar field models considered in this article. While we have seen in Sec.~\ref{sec:Models} that chameleons and symmetrons have $T^\nu_{\phantom{\nu}\nu}$-dependent VEVs and masses, we again follow Refs.~\cite{Burrage2018,Kading:2023mdk} and assume that the experiment takes place in a sufficiently small region in the center of a vacuum chamber, such that $\langle \varphi_\alpha \rangle$ and $m_\alpha$ are approximately constant. Such an assumption is needed in order to practically employ the path integral techniques presented in Refs.~\cite{Burrage2018,Kading2022x}, and it assures that the conformal factor in Eq.~(\ref{eq:FreeAtomAct}) leads to only a constant rescaling of the atom mass. Furthermore, we follow Refs.~\cite{Burrage2018,Kading:2023mdk,Fischer:2024gni} and assume that the screened scalar field had sufficient time to thermalise with the experimental setup, in particular, the vacuum chamber walls. Consequently, we must consider the $\chi_\alpha$-propagators with thermal corrections. For example, for the Feynman propagator we use (see, e.g., Ref.~\cite{Bellac:2011kqa})
\begin{eqnarray}
\label{eq:FeynProp}
    \Delta^{\rm F}_{\alpha}(x,y)
&=& -\mathrm{i} \int \frac{d^4k}{(2\pi)^4 } e^{ik\cdot (x-y)}\left[\frac{1}{k^2+m_\alpha^2-\mathrm{i}\epsilon} +2\pi \mathrm{i} f(|k^0|) \delta(k^2+m_\alpha^2) \right]~,
\end{eqnarray}
where 
\begin{eqnarray}
    f(k^0)&:=&\frac{1}{e^{ k^0/T}-1} 
\end{eqnarray}
is the Bose-Einstein distribution with the equilibrium temperature $T$. When $x=y$, i.e., in the case of a tadpole diagram, the thermal corrections for all four relevant types of propagators (Feynman, Dyson, positive and negative frequency Wightman) coincide and are ultraviolet-finite. They are of the form \cite{Burrage2018,Kading:2023mdk}
\begin{eqnarray}
\label{eq:TadPoltT}
\Delta^{(T\neq 0)}_{\alpha} 
&=& 
\int \frac{d^4k}{(2\pi)^3 }  f(|k^0|) \delta(k^2+m_\alpha^2)  \,=\,
\frac{T^2}{2\pi^2} \int_{m_\alpha/T}^\infty d\xi \frac{\sqrt{\xi^2-(\frac{m_\alpha}{T})^2}}{e^\xi -1}~.
\end{eqnarray}

In order to determine the coupling constants $\beta_\alpha$ and $\gamma_\alpha$ in Eq.~(\ref{eq:interact}) for the two considered screened scalar field models, we follow the discussion in Ref.~\cite{Burrage2018}. Ref.~\cite{Burrage2018} considers a general conformal factor of the form 
\begin{eqnarray}
  A_\alpha^2(\varphi_\alpha) = a + b \frac{\varphi_\alpha}{\mathcal{M}_\alpha} + c \frac{\varphi_\alpha^2}{\mathcal{M}_\alpha^2}~,   
\end{eqnarray}
and shows that 
\begin{eqnarray}
    \beta_\alpha 
    &=& A_\alpha(\langle \varphi_\alpha \rangle)
    \frac{M}{\mathcal{M}_\alpha}\Bigg[ \frac{b}{a} \left(1- \frac{b}{a}\frac{\langle \varphi_\alpha \rangle}{\mathcal{M}_\alpha} \right) 
    +
    2\frac{c}{a}\frac{\langle \varphi_\alpha \rangle}{\mathcal{M}_\alpha}
    \Bigg]
    ~,~~~ 
    \gamma_\alpha 
    \,=\, 
    2\frac{c}{a}A_\alpha^2(\langle \varphi_\alpha \rangle)
    \frac{M^2}{\mathcal{M}_\alpha^2}
\end{eqnarray}
must be fulfilled. From Eq.~(\ref{eq:ChamA}) we conclude that, for the chameleon, $a=1$ and $b=c=2$. Similarly, for the symmetron, Eq.~(\ref{eq:SymA}) lets us conclude that $a=c=1$ and $b=0$. Consequently, we find
\begin{eqnarray}
\label{eq:CouplingConst}
    \beta_C &\approx& 2\frac{M}{\mathcal{M}_C} \left( 1 +\frac{\langle \varphi_C \rangle}{\mathcal{M}_C}\right)
    ~,~ 
    \gamma_C \,\approx\, 4\frac{M^2}{\mathcal{M}_C^2}
    ~,~~~ 
    \beta_S \,\approx\, 2M \frac{\langle \varphi_S \rangle }{\mathcal{M}_S^2}
    ~,~ 
    \gamma_S \,\approx\, \frac{M}{\langle \varphi_S \rangle}\beta_S~,
\end{eqnarray}
where we have assumed $M, \langle \varphi_\alpha \rangle \ll \mathcal{M}_\alpha$ and only kept terms up to second order in $M/\mathcal{M}_\alpha$ and $\langle \varphi_\alpha \rangle/\mathcal{M}_\alpha$.

\begin{figure} [htbp]
\centering
    \subfloat[][]{\includegraphics[scale=0.3]{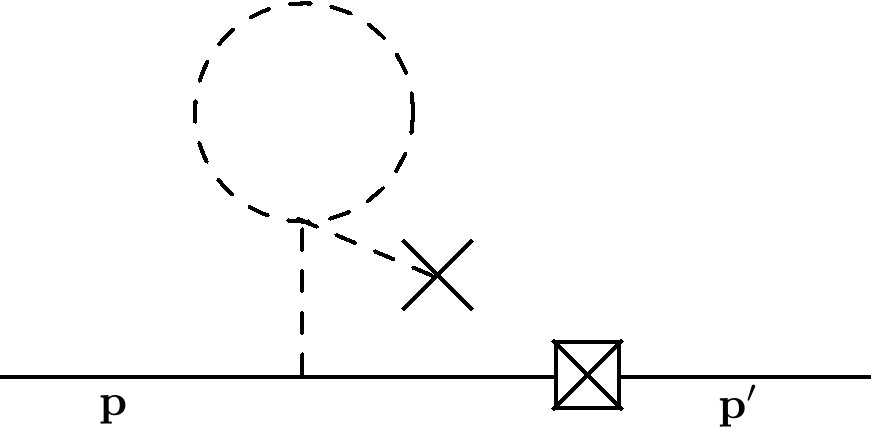}}
    \qquad
    \subfloat[][]{\includegraphics[scale=0.3]{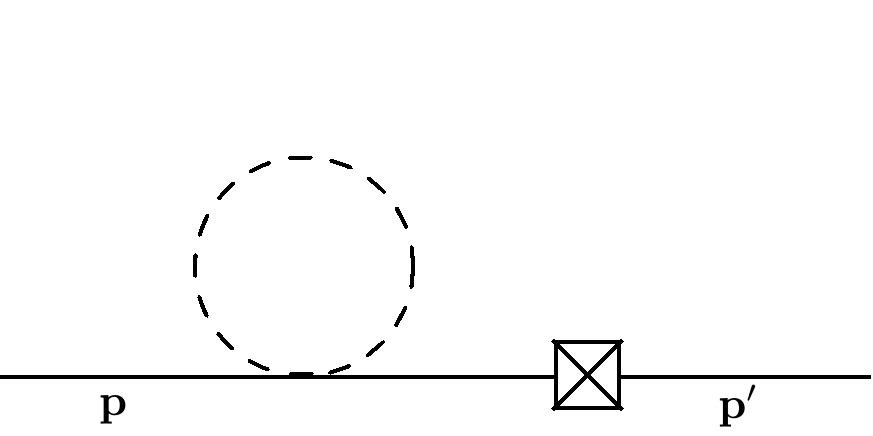}}
\caption{Diagrammatic depiction of the processes leading to the predicted frequency shifts, as given in Eqs.~(\ref{eq:DensCham}) and (\ref{eq:DensSym}), where a crossed box represents the reduced density matrix at the initial time, a solid line stands for an atom-propagator, and a dotted line is a propagator of a screened scalar field; note that the conjugated diagrams, i.e., those mirrored along a vertical line through the box and $\mathbf{p} \leftrightarrow \mathbf{p}'$, also contribute. (a) was taken from Ref.~\cite{Burrage2018} and (b) adapted from the same figure. (a): diagram leading to a frequency shift induced by $n=-4$ chameleons, where the X denotes the insertion of a VEV $\langle \varphi_C \rangle$; (b): diagram leading to a frequency shift induced by symmetrons}
    \label{fig:diagrams}
\end{figure}

Next, we follow the procedure that is in all detail outlined in Ref.~\cite{Kading:2023mdk}. This means that we want to use \cite{Kading2022x}
\begin{eqnarray}\label{eq:38Formula}
\rho_\alpha(\mathbf{p};\mathbf{p}' ;t)
&=& 
\lim_{\substack{x^{0(\prime)}\,\to\, t^{+}\\y^{0(\prime)}\,\to\, 0^-}}
\int \frac{d^3kd^3k'}{(2\pi)^6 4 E^{\phi,\alpha}_\mathbf{k} E^{\phi,\alpha}_{\mathbf{k}'}}
 \rho_\text{free}(\mathbf{k};\mathbf{k}';0) 
 \int_{\mathbf{x}\mathbf{x}'\mathbf{y}\mathbf{y}'} e^{-\mathrm{i}(\mathbf{p}\cdot\mathbf{x}-\mathbf{p}' \cdot\mathbf{x}')+\mathrm{i}(\mathbf{k}\cdot\mathbf{y}-\mathbf{k}'\cdot\mathbf{y}')}
\nonumber
\\
&&
\times 
\left( \partial_{x^0} - \mathrm{i}E^{\phi,\alpha}_{\mathbf{p}} \right)
\left( \partial_{x^{0\prime}} + \mathrm{i}E^{\phi,\alpha}_{\mathbf{p}'} \right)
\left( \partial_{y^0} + \mathrm{i}E^{\phi,\alpha}_{\mathbf{k}} \right)
\left( \partial_{y^{0\prime}} - \mathrm{i}E^{\phi,\alpha}_{\mathbf{k}'} \right)
\nonumber
\\
&&
\times 
\int\mathcal{D}\phi^{+}\mathcal{D}\phi^{-} e^{\mathrm{i}\{S_{\phi}[\phi^+]-S_{\phi}[\phi^-]\}}\phi^+(x)\phi^-(x')\mathcal{F}_\alpha[\phi^{+};\phi^{-};t]\phi^{+}(y)\phi^{-}(y')~,
\end{eqnarray}
where $\phi^+$ and $\phi^-$ denote scalar fields associated with the positive and negative branches of the Schwinger-Keldysh closed time path \cite{Schwinger,Keldysh},
in order to find an expression like Eq.~(\ref{eq:NonUniEv}). The Feynman-Vernon influence functional $\mathcal{F}_\alpha[\phi^{+};\phi^{-};t]$ \cite{Feynman} describes the effects of the screened scalar fields on the open dynamics of the probe atom, and is given as a trace over the $\chi_\alpha$-degrees of freedom \cite{Kading2022x}: 
\begin{eqnarray}
\mathcal{F}_\alpha[\phi^{+};\phi^{-};t] &=& \left\langle \exp\big\{ \mathrm{i}\big( S_{\text{self}}[\chi_\alpha^+;t] - S_{\text{self}}[\chi_\alpha^-;t] + S_\text{int}[\phi^+,\chi_\alpha^+;t] - S_\text{int}[\phi^-,\chi_\alpha^-;t] \big) \big\} \right\rangle_{\chi_\alpha}~.
\end{eqnarray}
For practical purposes, we can only compute it perturbatively. From Ref.~\cite{Burrage2018}, we already know which terms of the Feynman-Vernon influence functional will give the leading effect when discussing chameleons. However, note that this requires us to only consider cases for which $\lambda_C \gg \beta_C$. Similarly, since the symmetron has basically the same matter coupling as the environment-dependent dilaton, we can use Ref.~\cite{Kading:2023mdk} in order to determine the leading term for this model. Consequently, we only need to consider:
\begin{eqnarray}
\label{eq:FVICham}
    \mathcal{F}_C[\phi^{+};\phi^{-};t] &=& 1 - \frac{\lambda_C\beta_C}{8}M \langle \varphi_C \rangle \sum_{\kappa,\nu=\pm} \kappa\nu\int_{zz'} \Delta^{\rm F}_{C}(z,z) \Delta^{\kappa\nu}_{C}(z,z') 
    [\phi^\nu(z')]^2
    +\dots~,
    \\
    \label{eq:FVISym}
    \mathcal{F}_S[\phi^{+};\phi^{-};t] &=&  1 - \mathrm{i} \frac{\gamma_S}{4} \sum_{\kappa=\pm} \kappa \int_z \Delta^{\rm F}_{S}(z,z)[\phi^\kappa(z)]^2  +  \ldots~.
\end{eqnarray}
Note that $\Delta^{++}$ is a Feynman propagator, $\Delta^{--}$ is a Dyson propagator, and $\Delta^{+-}$ and $\Delta^{-+}$ are Wightman propagators; see, e.g., Ref.~\cite{Burrage2018}. Substituting Eqs.~(\ref{eq:FVICham}) and (\ref{eq:FVISym}) into Eq.~(\ref{eq:38Formula}), and computing the path integrals over $\phi^+$ and $\phi^-$, we find
\begin{eqnarray}\label{eq:DensCham}
\rho_C(\mathbf{p};\mathbf{p}' ;t)
&\approx& 
\lim_{\substack{x^{0(\prime)}\,\to\, t^{+}\\y^{0(\prime)}\,\to\, 0^-}}
\int \frac{d^3kd^3k'}{(2\pi)^6 4 E^{\phi,C}_\mathbf{k} E^{\phi,C}_{\mathbf{k}'}}
 \rho_\text{free}(\mathbf{k};\mathbf{k}';0) 
 \int_{\mathbf{x}\mathbf{x}'\mathbf{y}\mathbf{y}'} e^{-\mathrm{i}(\mathbf{p}\cdot\mathbf{x}-\mathbf{p}' \cdot\mathbf{x}')+\mathrm{i}(\mathbf{k}\cdot\mathbf{y}-\mathbf{k}'\cdot\mathbf{y}')}
\nonumber
\\
&&
\times 
\left( \partial_{x^0} - \mathrm{i}E^{\phi,C}_{\mathbf{p}} \right)
\left( \partial_{x^{0\prime}} + \mathrm{i}E^{\phi,C}_{\mathbf{p}'} \right)
\left( \partial_{y^0} + \mathrm{i}E^{\phi,C}_{\mathbf{k}} \right)
\left( \partial_{y^{0\prime}} - \mathrm{i}E^{\phi,C}_{\mathbf{k}'} \right)
\nonumber
\\
&&
\times 
\Big\{
D^\mathrm{++}(x,y)D^\mathrm{--}(x',y') 
\nonumber
\\
&&
~~~~~~
- \frac{\lambda_C\beta_C}{4}M \langle \varphi_C \rangle \sum_{\kappa} \kappa\int_{zz'} \Delta^{\rm F}_{C}(z,z) 
\nonumber
\\
&&
~~~~~~~~~~~~~
\times
    \left[ \Delta^{\kappa+}_{C}(z,z')D^\mathrm{++}(x,z)D^\mathrm{++}(z,y)D^\mathrm{--}(x',y') - (x,y \longleftrightarrow x',y')^\ast \right]
    \Big\}
    ~,~~~
\\
\label{eq:DensSym}
\rho_S(\mathbf{p};\mathbf{p}' ;t)
&\approx& 
\lim_{\substack{x^{0(\prime)}\,\to\, t^{+}\\y^{0(\prime)}\,\to\, 0^-}}
\int \frac{d^3kd^3k'}{(2\pi)^6 4 E^{\phi,S}_\mathbf{k} E^{\phi,S}_{\mathbf{k}'}}
 \rho_\text{free}(\mathbf{k};\mathbf{k}';0) 
 \int_{\mathbf{x}\mathbf{x}'\mathbf{y}\mathbf{y}'} e^{-\mathrm{i}(\mathbf{p}\cdot\mathbf{x}-\mathbf{p}' \cdot\mathbf{x}')+\mathrm{i}(\mathbf{k}\cdot\mathbf{y}-\mathbf{k}'\cdot\mathbf{y}')}
\nonumber
\\
&&
\times 
\left( \partial_{x^0} - \mathrm{i}E^{\phi,S}_{\mathbf{p}} \right)
\left( \partial_{x^{0\prime}} + \mathrm{i}E^{\phi,S}_{\mathbf{p}'} \right)
\left( \partial_{y^0} + \mathrm{i}E^{\phi,S}_{\mathbf{k}} \right)
\left( \partial_{y^{0\prime}} - \mathrm{i}E^{\phi,S}_{\mathbf{k}'} \right)
\nonumber
\\
&&
\times 
\Big\{
D^\mathrm{++}(x,y)D^\mathrm{--}(x',y')  
\nonumber
\\
&&
~~~~~~
- \mathrm{i} \frac{\gamma_S}{2}  \int_z \Delta^{\rm F}_{S}(z,z)\left[ D^\mathrm{++}(x,z)D^\mathrm{++}(z,y)D^\mathrm{--}(x',y') - (x,y \longleftrightarrow x',y')^\ast \right] 
\Big\}
~,~~~
\end{eqnarray}
where $D$ denotes $\phi$-propagators, we have dropped disconnected bubble diagrams, and, for notational convenience, we have omitted the $+ \ldots$ . The first terms in the curly brackets of Eqs.~(\ref{eq:DensCham}) and (\ref{eq:DensSym}) are the usual unitary evolution terms but with atom masses rescaled by factors of $A_\alpha^2(\langle \varphi_\alpha \rangle)$, while the last terms are inducing frequency shifts $ \Delta u_\alpha(\mathbf{p},\mathbf{p}')$. We can illustrate the relevant physical processes that induce the frequency shifts by translating their propagator expressions into Feynman diagrams; see Fig.~\ref{fig:diagrams}.
After evaluating the remaining integrals in Eqs.~(\ref{eq:DensCham}) and (\ref{eq:DensSym}), we finally obtain
\begin{eqnarray}
\rho_C(\mathbf{p};\mathbf{p}' ;t)
&\approx& 
\rho_\text{free}(\mathbf{p}; \mathbf{p'};0) e^{-\mathrm{i}(E^{\phi,C}_{\mathbf{p}} - E^{\phi,C}_{\mathbf{p'}}) t}
\nonumber
\\
&&
\times
  \Bigg\{
1 - \frac{\mathrm{i}\lambda_C\beta_C}{4} \frac{M \langle \varphi_C \rangle}{m_C^2} \Delta^{\rm F}_{C}(0)\left( \frac{1}{E^{\phi,C}_{\mathbf{p}}} - \frac{1}{E^{\phi,C}_{\mathbf{p'}}} \right) 
\left[\frac{\sin(m_C t)}{m_C}-t\right]
\Bigg\}
    ~,
\\
\rho_S(\mathbf{p};\mathbf{p}' ;t)
&\approx& 
\rho_\text{free}(\mathbf{p}; \mathbf{p'};0) e^{-\mathrm{i}(E^{\phi,S}_{\mathbf{p}} - E^{\phi,S}_{\mathbf{p'}}) t}
\Bigg[
1 - \frac{\mathrm{i} \gamma_S}{4} \Delta^{\rm F}_{S}(0)\left( \frac{1}{E^{\phi,S}_{\mathbf{p}}} - \frac{1}{E^{\phi,S}_{\mathbf{p'}}} \right) t 
\Bigg]
~,
\end{eqnarray}
where $\Delta^{\rm F}_{\alpha}(0)$ denotes a tadpole propagator. Subsequently, we follow Ref.~\cite{Burrage2018} and add counter terms 
\begin{eqnarray}
\delta S_{\text{self}}[\chi_\alpha] &:=& \frac{\lambda_\alpha \langle \varphi_\alpha \rangle}{2} \int_{x}  \Delta^{F(T=0)}_\alpha \chi_\alpha~,
~~~
\delta S_{\text{int}}[\phi, \chi_\alpha] \,:=\,    \ \frac{\gamma_\alpha}{4} \int_{x}  \Delta^{F(T=0)}_\alpha  \phi^2  
\end{eqnarray}
to the actions in Eqs.~(\ref{eq:SelfIntact}) and (\ref{eq:interact}) in order to deal with the divergences originating from the tadpoles, where $\Delta^{F(T=0)}_\alpha$ is the Feynman propagator given in Eq.~(\ref{eq:FeynProp}) for $T=0$ and $x=y$. As a consequence, we are left with
\begin{eqnarray}
\label{eq:rhoCaftR}
\rho_C(\mathbf{p};\mathbf{p}' ;t)
&\approx& 
\rho_\text{free}(\mathbf{p}; \mathbf{p'};0) e^{-\mathrm{i}(E^{\phi,C}_{\mathbf{p}} - E^{\phi,C}_{\mathbf{p'}}) t}
\nonumber
\\
&&
\times
  \Bigg\{
1 - \frac{\mathrm{i}\lambda_C\beta_C}{4} \frac{M \langle \varphi_C \rangle}{m_C^2} \Delta^{(T\neq 0)}_{C}\left( \frac{1}{E^{\phi,C}_{\mathbf{p}}} - \frac{1}{E^{\phi,C}_{\mathbf{p'}}} \right) 
\left[\frac{\sin(m_C t)}{m_C}-t\right]
\Bigg\}
    ~,
\\
\label{eq:rhoSaftR}
\rho_S(\mathbf{p};\mathbf{p}' ;t)
&\approx& 
\rho_\text{free}(\mathbf{p}; \mathbf{p'};0) e^{-\mathrm{i}(E^{\phi,S}_{\mathbf{p}} - E^{\phi,S}_{\mathbf{p'}}) t}
\Bigg[
1 - \frac{\mathrm{i} \gamma_S}{4} \Delta^{(T\neq 0)}_{S}\left( \frac{1}{E^{\phi,S}_{\mathbf{p}}} - \frac{1}{E^{\phi,S}_{\mathbf{p'}}} \right) t 
\Bigg]
\end{eqnarray}
with $\Delta^{(T\neq 0)}_{\alpha}$ given in Eq.~(\ref{eq:TadPoltT}). Next, we expand the energies with the rescaled masses in terms of $\langle \varphi_\alpha \rangle/\mathcal{M}_\alpha$; substitute Eq.~(\ref{eq:CouplingConst}) into Eqs.~(\ref{eq:rhoCaftR}) and (\ref{eq:rhoSaftR}); keep only terms up to second order in $\lambda_C$, $M/\mathcal{M}_\alpha$ and $\langle \varphi_\alpha \rangle/\mathcal{M}_\alpha$; and interpret the last line of Eq.~(\ref{eq:rhoCaftR}) and the term in the square brackets in Eq.~(\ref{eq:rhoSaftR}) as expansions of exponential functions. In this way, we find
\begin{eqnarray}
\rho_C(\mathbf{p};\mathbf{p}' ;t)
&\approx& 
\rho_\text{free}(\mathbf{p}; \mathbf{p'};0) 
e^{-\mathrm{i}[E^{\phi}_{\mathbf{p}} - E^{\phi}_{\mathbf{p'}} + \Delta u_C(\mathbf{p},\mathbf{p}')] t}
~,
\\
\rho_S(\mathbf{p};\mathbf{p}' ;t)
&\approx& 
\rho_\text{free}(\mathbf{p}; \mathbf{p'};0) e^{-\mathrm{i}[E^{\phi}_{\mathbf{p}} - E^{\phi}_{\mathbf{p'}} + \Delta u_S(\mathbf{p},\mathbf{p}')] t}~,
\end{eqnarray}
where, by comparison with Eq.~(\ref{eq:NonUniEv}), the frequency shifts induced by $n=-4$ chameleons and symmetrons between two atomic states with momenta $\mathbf{p}$ and $\mathbf{p}'$ are
\begin{eqnarray}
    \Delta u_C(\mathbf{p},\mathbf{p}') &\approx& 
    \frac{M^2 \langle \varphi_C\rangle}{\mathcal{M}_{C}}
\left\{  1 
-
\frac{\lambda_C}{2 m_C^2}\Delta^{(T\neq 0)}_{C} [1-\text{sinc}(m_C t)]\right\}
 \left( \frac{1}{E_\mathbf{p}^\phi} - \frac{1}{E_{\mathbf{p}'}^\phi} \right)
    ~,
    \\
    \Delta u_S(\mathbf{p},\mathbf{p}') &\approx& \frac{M^2}{2\mathcal{M}_{S}^2}
\left(  \langle \varphi_S\rangle^2 
+
\Delta^{(T\neq 0)}_{S} \right)
 \left( \frac{1}{E_\mathbf{p}^\phi} - \frac{1}{E_{\mathbf{p}'}^\phi} \right)
\end{eqnarray}
with $\text{sinc}(m_C t) = \sin(m_C t)/m_C t$.

Next, we consider the non-relativistic case $|\mathbf{p}|, |\mathbf{p}'| \ll M $ and introduce $v := \frac{||\mathbf{p}|-|\mathbf{p}'||}{M}$ as the speed difference between two atomic states, such that the frequency shifts become
\begin{eqnarray}
\label{eq:ChamShift}
    \Delta u_C &\approx& 
    \frac{M \langle \varphi_C\rangle}{2\mathcal{M}_{C}}
\left\{  1 
-
\frac{\lambda_C}{2 m_C^2}\Delta^{(T\neq 0)}_{C} [1-\text{sinc}(m_C t)]\right\}
v^2
    ~,
    \\
\label{eq:SymShift}
    \Delta u_S &\approx& \frac{M}{4\mathcal{M}_{S}^2} \left(  \langle \varphi_S\rangle^2 
+
\Delta^{(T\neq 0)}_{S} \right)v^2
    ~.
\end{eqnarray}
Note that the chameleon-induced frequency shift that we have derived in Eq.~(\ref{eq:ChamShift}) has some similarities with the one found in Ref.~\cite{Burrage2018}. However, they are clearly not identical. This stems from the fact that Ref.~\cite{Burrage2018} reads off the frequency shift directly from a quantum master equation, which is an incorrect procedure because the shift is $t$-dependent and cannot be extracted without actually solving the master equation. Since the method presented in Ref.~\cite{Kading2022x} allows us to directly compute the reduced density matrix elements, we were able to find the correct frequency shift in this article. In addition, in contrast to the result in Ref.~\cite{Burrage2018}, Eq.~(\ref{eq:ChamShift}) allows for arbitrary temperatures and still has its explicit time dependence. This renders it more general.

Next, we check whether the frequency shifts in Eqs.~(\ref{eq:ChamShift}) and (\ref{eq:SymShift}) could actually lead to new constraints on the considered models. For this, we use the same experimental parameters as in Ref.~\cite{Kading:2023mdk}, i.e., we use a vacuum chamber with radius $L = 10$ cm \cite{Burrage:2015lya,Burrage:2014oza}; assume a speed difference $v=50$ mm $\text{s}^{-1}$ \cite{bcca-m22}; and state that the smallest frequency shifts that can currently be measured in atom interferometry experiments are $\Delta u_\text{min} \approx 10^{-8}$ Hz \cite{bcca-m22,eymkl15}. In order to avoid any non-negligible effects of the experimental setup on the screened scalar fields' VEVs and masses, we restrict our discussion to model parameters for which $L \gg 1/m_\alpha \gg r_\text{nuc}$ is fulfilled, where $r_\text{nuc}$ is the radius of our probe atom's nucleus. Note that we take $r_\text{nuc}$ instead of the atomic radius since the nucleus is by far the most dense part of an atom and is consequently expected to have the largest impact on the screened scalar fields. Furthermore, in order to be consistent with our previous assumptions, we must restrict our discussion to the cases $\lambda_\alpha \ll 1$, $\langle\varphi_\alpha\rangle \ll \mathcal{M}_\alpha$, and $M \ll \mathcal{M}_\alpha$. For the chameleon, we consider two limiting cases, $m_C t \to 0$ and $m_C t \to \infty$, such that we can make use of $\lim_{x \to0} \text{sinc}(x) = 1$ and $\lim_{x \to\infty} \text{sinc}(x) = 0$. The symmetron-induced frequency shift is discussed only for parts of the parameter space in which the fifth force is unscreened.

\begin{figure} [htbp]
\centering
    \subfloat[][]{\includegraphics[scale=0.45]{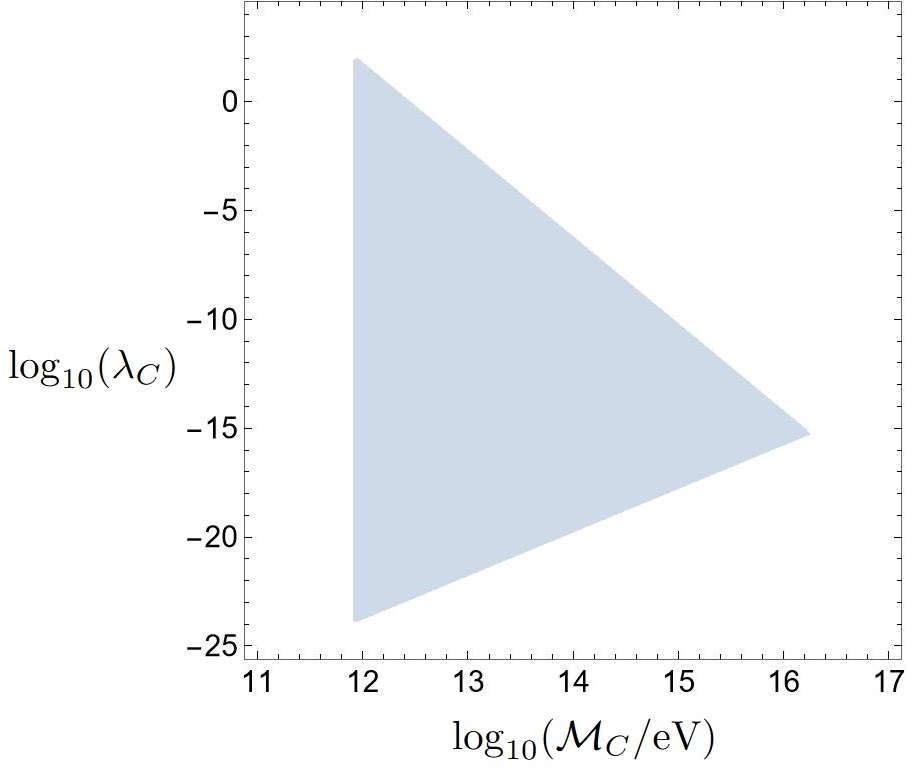}}
    
    \subfloat[][]{\includegraphics[scale=0.45]{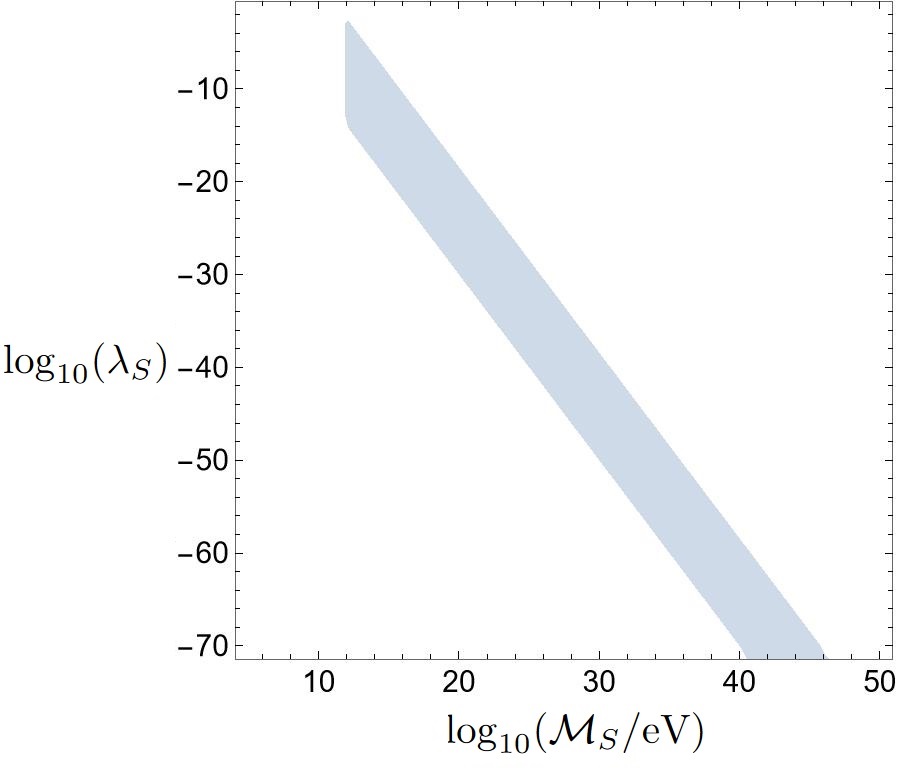}}
    \qquad
    \subfloat[][]{\includegraphics[scale=0.45]{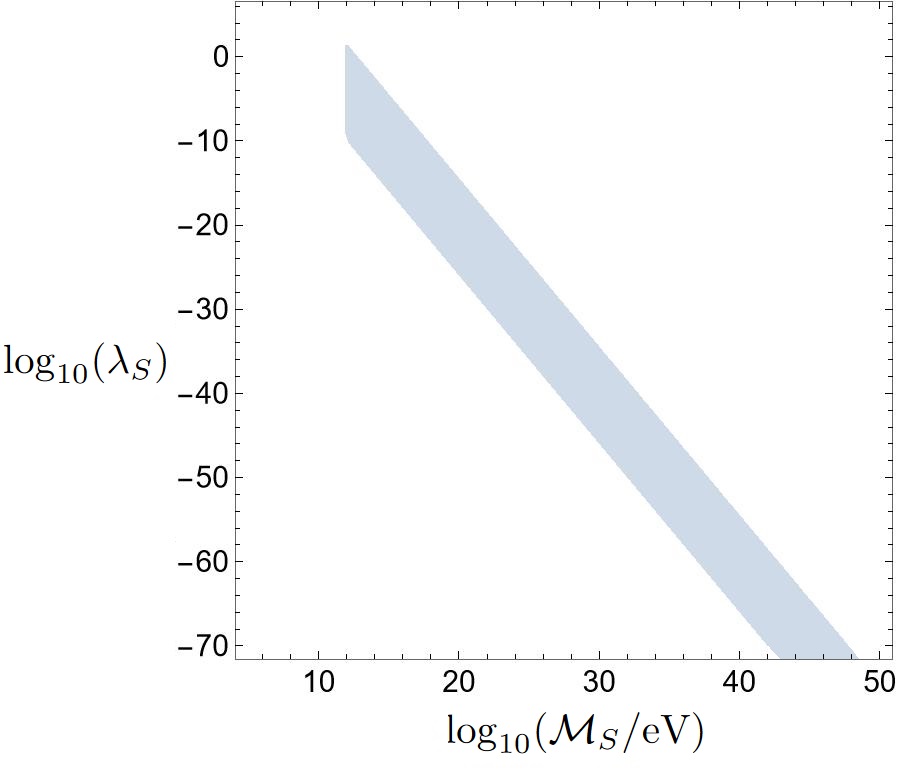}}
\caption{Predicted constraints for the two screened scalar field models when comparing the frequency shifts induced in two atom interferometry experiments with either Rb-87 or Li-7; (a): $n=-4$ chameleon constraints in a parameter space region that has already been covered by other experiments \cite{Burrage:2017qrf}; (b) and (c): symmetron constraints for $\mu = 10^4$ eV and $\mu = 10^6$ eV, respectively; the potential constraints for the symmetron are on parts of the parameter space that have never been constrained by any other experiment before; see Ref.~\cite{Fischer:2024eic} for the most recent summary of symmetron constraints. On the abscissae of (b) and (c), the distances to any constraints given in Ref.~\cite{Fischer:2024eic} are larger than $10$ orders of magnitude. }
    \label{fig:plots}
\end{figure}

As was pointed out in Ref.~\cite{Burrage2018} and implemented in Ref.~\cite{Kading:2023mdk}, frequency shifts can only be quantified if one measurement is compared to another measurement under different experimental conditions. This means that we actually need to predict frequency shifts for two distinct experimental setups and look at the difference of those. If we predict a difference in frequency shifts that is at least $\Delta u_\text{min}$ but it is not observed in actual experiments, then this constrains the parts of the screened scalar field parameter spaces for which we have made our predictions. At first, we follow the path of Ref.~\cite{Kading:2023mdk}, i.e., we take a rubidium-87 atom as the probe particle, and predict frequency shifts for different temperatures and gas pressures in the vacuum chamber. However, in this way, we do not find any new constraints on either of the two screened scalar field models considered in this article. Therefore, we employ a different approach and actually compare two experiments at the same temperature, $T = 300$ K, and with the same residual $\text{H}_2$ gas pressure, $P = 9.6 \times 10^{-10}$ mbar \cite{Sabulsky:2018jma}, but with two different atom species. Besides rubidium-87, lithium-7 is used in atom interferometry \cite{Lepoutre:2010zz}. Comparing the predicted frequency shifts for experiments with those two types of atoms, i.e., we search for parameters that fulfill
\begin{eqnarray}
    |\Delta u_\alpha(M_\text{Rb-87}) - \Delta u_\alpha(M_\text{Li-7})| \geq \Delta u_\text{min}~,
\end{eqnarray}
we expect to find constraints on the screened scalar field models as shown in Fig.~\ref{fig:plots}. 
In actual experiments, we would of course have to take into account that there is a natural difference in the unitary evolutions for both types of atoms due to the different atomic masses. Interestingly, our predictions for the chameleon are not noticeably affected by choosing either of the two cases $m_C t \to 0$ or $m_C t \to \infty$, which implies that the thermal correction is subleading. Unfortunately, for the $n=-4$ chameleon, our results show that we do not have to expect induced frequency shifts to lead to novel constraints. However, this does not necessarily come as a surprise since this model is already tightly constrained. Though, our predictions for symmetrons indicate that open quantum dynamical effects have the potential to constrain parts of the model parameter space that have never been reached by any experiment before. Quite interestingly, frequency shifts seem to give particularly tight constraints for values of $\mu \approx 10^4$ eV and above, a regime that can currently only be reached by a few experiments; see Ref.~\cite{Fischer:2024eic}.


\section{Conclusions}
\label{sec:Conclusion}

Light scalar fields appear in many contexts throughout modern physics. Consequently, they are actively searched for in many experiments. Screened scalar fields are some particularly interesting light scalar models since they dynamically circumvent current Solar System constraints on the fifth forces they are expected to cause. Despite enormous experimental efforts, many of these models are, as of yet, not fully excluded. This fuels the demands for new experimental ideas. 

In this article, we have complemented the studies of Refs.~\cite{Burrage2018,Kading:2023mdk} by deriving and discussing frequency shifts induced in atom interferometry experiments due to interactions with two well-known screened scalar field models, $n=-4$ chameleons and symmetrons. We have shown that the original prediction of Ref.~\cite{Burrage2018} was not fully correct. Employing more recently developed methods from Ref.~\cite{Kading2022x}, we were able to properly derive expressions for frequency shifts as corrections to the unitary dynamics of an open scalar field system that acted as a proxy for a cold atom. While a scalar field representing an atom is a rather rough approximation, we have argued why this can still give us a first estimation on whether open quantum dynamical effects induced by chameleons or symmetrons can lead to constraints on these models. Using more sophisticated computational methods that are currently still in development, we will in the future be able to confirm or amend the results from the present article.    

While we have shown that frequency shifts can likely only be used to investigate regions of the $n=-4$ chameleon parameter space that have already been covered by other experiments, they could still be useful in order to confirm such existing constraints. However, much more interesting is our conclusion that frequency shifts have the potential to tightly constrain parts of the symmetron parameter space that are currently out of reach for any other experiment. In particular, our results indicate that frequency shifts can be powerful tools when studying symmetrons with tachyonic masses above the keV scale. 


\begin{acknowledgments}
This research was funded in whole or in part by the Austrian Science Fund (FWF) 
[10.55776/\allowbreak PAT8564023], and is based upon work from COST Action COSMIC WISPers CA21106, supported by COST (European Cooperation in Science and Technology). For open access purposes, the author has applied a CC BY public copyright license to any author accepted manuscript version arising from this submission. The authors acknowledge TU Wien Bibliothek for financial support through its Open Access Funding Programme.
\end{acknowledgments}


\bibliography{Bib}

\providecommand{\href}[2]{#2}\begingroup\raggedright\begin{thebibliography}{10}

\bibitem{Fujii2003}
{Fujii, Yasunori and Maeda, Kei-ichi}, \emph{The Scalar-Tensor Theory of
  Gravitation}, Cambridge Monographs on Mathematical Physics. Cambridge
  University Press, 2003,
  \href{https://doi.org/10.1017/CBO9780511535093}{10.1017/CBO9780511535093}.

\bibitem{Clifton2011}
T.~Clifton, P.~G. Ferreira, A.~Padilla and C.~Skordis, \emph{Modified gravity
  and cosmology},
  \href{https://doi.org/https://doi.org/10.1016/j.physrep.2012.01.001}{\emph{Physics
  Reports} {\bfseries 513} (2012) 1}.

\bibitem{Joyce2014}
A.~Joyce, B.~Jain, J.~Khoury and M.~Trodden, \emph{{Beyond the Cosmological
  Standard Model}},
  \href{https://doi.org/10.1016/j.physrep.2014.12.002}{\emph{Phys. Rept.}
  {\bfseries 568} (2015) 1} [\href{https://arxiv.org/abs/1407.0059}{{\ttfamily
  1407.0059}}].

\bibitem{Dickey1994}
J.~O. Dickey, P.~L. Bender, J.~E. Faller, X.~X. Newhall, R.~L. Ricklefs, J.~G.
  Ries et~al., \emph{{Lunar Laser Ranging: A Continuing Legacy of the Apollo
  Program}}, \href{https://doi.org/10.1126/science.265.5171.482}{\emph{Science}
  {\bfseries 265} (1994) 482}.

\bibitem{Adelberger2003}
E.~Adelberger, B.~Heckel and A.~Nelson, \emph{{Tests of the Gravitational
  Inverse-Square Law}},
  \href{https://doi.org/10.1146/annurev.nucl.53.041002.110503}{\emph{Annual
  Review of Nuclear and Particle Science} {\bfseries 53} (2003) 77}.

\bibitem{Kapner2007}
D.~J. Kapner, T.~S. Cook, E.~G. Adelberger, J.~H. Gundlach, B.~R. Heckel, C.~D.
  Hoyle et~al., \emph{{Tests of the Gravitational Inverse-Square Law below the
  Dark-Energy Length Scale}},
  \href{https://doi.org/10.1103/PhysRevLett.98.021101}{\emph{Phys. Rev. Lett.}
  {\bfseries 98} (2007) 021101}.

\bibitem{Khoury2003}
J.~Khoury and A.~Weltman, \emph{{Chameleon cosmology}},
  \href{https://doi.org/10.1103/PhysRevD.69.044026}{\emph{Phys. Rev. D}
  {\bfseries 69} (2004) 044026}
  [\href{https://arxiv.org/abs/astro-ph/0309411}{{\ttfamily
  astro-ph/0309411}}].

\bibitem{Khoury20032}
J.~Khoury and A.~Weltman, \emph{{Chameleon fields: Awaiting surprises for tests
  of gravity in space}},
  \href{https://doi.org/10.1103/PhysRevLett.93.171104}{\emph{Phys. Rev. Lett.}
  {\bfseries 93} (2004) 171104}
  [\href{https://arxiv.org/abs/astro-ph/0309300}{{\ttfamily
  astro-ph/0309300}}].

\bibitem{Dehnen1992}
H.~Dehnen, H.~Frommert and F.~Ghaboussi, \emph{{Higgs field and a new scalar -
  tensor theory of gravity}},
  \href{https://doi.org/10.1007/BF00674344}{\emph{Int. J. Theor. Phys.}
  {\bfseries 31} (1992) 109}.

\bibitem{Gessner1992}
E.~Gessner, \emph{{A new scalar tensor theory for gravity and the flat rotation
  curves of spiral galaxies}},
  \href{https://doi.org/10.1007/BF00645239}{\emph{Astrophys. Space Sci.}
  {\bfseries 196} (1992) 29}.

\bibitem{Damour1994}
T.~Damour and A.~M. Polyakov, \emph{{The String dilaton and a least coupling
  principle}}, \href{https://doi.org/10.1016/0550-3213(94)90143-0}{\emph{Nucl.
  Phys. B} {\bfseries 423} (1994) 532}
  [\href{https://arxiv.org/abs/hep-th/9401069}{{\ttfamily hep-th/9401069}}].

\bibitem{Pietroni2005}
M.~Pietroni, \emph{Dark energy condensation},
  \href{https://doi.org/10.1103/PhysRevD.72.043535}{\emph{Phys. Rev. D}
  {\bfseries 72} (2005) 043535}.

\bibitem{Olive2008}
K.~A. Olive and M.~Pospelov, \emph{Environmental dependence of masses and
  coupling constants},
  \href{https://doi.org/10.1103/PhysRevD.77.043524}{\emph{Phys. Rev. D}
  {\bfseries 77} (2008) 043524}.

\bibitem{Brax2010}
P.~Brax, C.~van~de Bruck, A.-C. Davis and D.~Shaw, \emph{Dilaton and modified
  gravity}, \href{https://doi.org/10.1103/PhysRevD.82.063519}{\emph{Phys. Rev.
  D} {\bfseries 82} (2010) 063519}.

\bibitem{Hinterbichler2010}
K.~Hinterbichler and J.~Khoury, \emph{{Symmetron Fields: Screening Long-Range
  Forces Through Local Symmetry Restoration}},
  \href{https://doi.org/10.1103/PhysRevLett.104.231301}{\emph{Phys. Rev. Lett.}
  {\bfseries 104} (2010) 231301}
  [\href{https://arxiv.org/abs/1001.4525}{{\ttfamily 1001.4525}}].

\bibitem{Hinterbichler2011}
K.~Hinterbichler, J.~Khoury, A.~Levy and A.~Matas, \emph{{Symmetron
  Cosmology}}, \href{https://doi.org/10.1103/PhysRevD.84.103521}{\emph{Phys.
  Rev. D} {\bfseries 84} (2011) 103521}
  [\href{https://arxiv.org/abs/1107.2112}{{\ttfamily 1107.2112}}].

\bibitem{Fischer:2024eic}
H.~Fischer, C.~K\"ading and M.~Pitschmann, \emph{{Screened Scalar Fields in the
  Laboratory and the Solar System}},
  \href{https://doi.org/10.3390/universe10070297}{\emph{Universe} {\bfseries
  10} (2024) 297} [\href{https://arxiv.org/abs/2405.14638}{{\ttfamily
  2405.14638}}].

\bibitem{Burrage:2017qrf}
C.~Burrage and J.~Sakstein, \emph{{Tests of Chameleon Gravity}},
  \href{https://doi.org/10.1007/s41114-018-0011-x}{\emph{Living Rev. Rel.}
  {\bfseries 21} (2018) 1} [\href{https://arxiv.org/abs/1709.09071}{{\ttfamily
  1709.09071}}].

\bibitem{Faulkner:2006ub}
T.~Faulkner, M.~Tegmark, E.~F. Bunn and Y.~Mao, \emph{{Constraining f(R)
  Gravity as a Scalar Tensor Theory}},
  \href{https://doi.org/10.1103/PhysRevD.76.063505}{\emph{Phys. Rev. D}
  {\bfseries 76} (2007) 063505}
  [\href{https://arxiv.org/abs/astro-ph/0612569}{{\ttfamily
  astro-ph/0612569}}].

\bibitem{Burrage:2016yjm}
C.~Burrage, E.~J. Copeland and P.~Millington, \emph{{Radial acceleration
  relation from symmetron fifth forces}},
  \href{https://doi.org/10.1103/PhysRevD.95.064050}{\emph{Phys. Rev. D}
  {\bfseries 95} (2017) 064050}
  [\href{https://arxiv.org/abs/1610.07529}{{\ttfamily 1610.07529}}].

\bibitem{OHare:2018ayv}
C.~A.~J. O'Hare and C.~Burrage, \emph{{Stellar kinematics from the symmetron
  fifth force in the Milky Way disk}},
  \href{https://doi.org/10.1103/PhysRevD.98.064019}{\emph{Phys. Rev. D}
  {\bfseries 98} (2018) 064019}
  [\href{https://arxiv.org/abs/1805.05226}{{\ttfamily 1805.05226}}].

\bibitem{Burrage:2018zuj}
C.~Burrage, E.~J. Copeland, C.~K\"ading and P.~Millington, \emph{{Symmetron
  scalar fields: Modified gravity, dark matter, or both?}},
  \href{https://doi.org/10.1103/PhysRevD.99.043539}{\emph{Phys. Rev. D}
  {\bfseries 99} (2019) 043539}
  [\href{https://arxiv.org/abs/1811.12301}{{\ttfamily 1811.12301}}].

\bibitem{Kading:2023hdb}
C.~K\"ading, \emph{{Lensing with Generalized Symmetrons}},
  \href{https://doi.org/10.3390/astronomy2020009}{\emph{Astronomy} {\bfseries
  2} (2023) 128} [\href{https://arxiv.org/abs/2304.05875}{{\ttfamily
  2304.05875}}].

\bibitem{Burrage:2014oza}
C.~Burrage, E.~J. Copeland and E.~A. Hinds, \emph{{Probing Dark Energy with
  Atom Interferometry}},
  \href{https://doi.org/10.1088/1475-7516/2015/03/042}{\emph{JCAP} {\bfseries
  03} (2015) 042} [\href{https://arxiv.org/abs/1408.1409}{{\ttfamily
  1408.1409}}].

\bibitem{Hamilton:2015zga}
P.~Hamilton, M.~Jaffe, P.~Haslinger, Q.~Simmons, H.~M\"uller and J.~Khoury,
  \emph{{Atom-interferometry constraints on dark energy}},
  \href{https://doi.org/10.1126/science.aaa8883}{\emph{Science} {\bfseries 349}
  (2015) 849} [\href{https://arxiv.org/abs/1502.03888}{{\ttfamily
  1502.03888}}].

\bibitem{Lemmel:2015kwa}
H.~Lemmel, P.~Brax, A.~N. Ivanov, T.~Jenke, G.~Pignol, M.~Pitschmann et~al.,
  \emph{{Neutron Interferometry constrains dark energy chameleon fields}},
  \href{https://doi.org/10.1016/j.physletb.2015.02.063}{\emph{Phys. Lett. B}
  {\bfseries 743} (2015) 310}
  [\href{https://arxiv.org/abs/1502.06023}{{\ttfamily 1502.06023}}].

\bibitem{Burrage:2015lya}
C.~Burrage and E.~J. Copeland, \emph{{Using Atom Interferometry to Detect Dark
  Energy}}, \href{https://doi.org/10.1080/00107514.2015.1060058}{\emph{Contemp.
  Phys.} {\bfseries 57} (2016) 164}
  [\href{https://arxiv.org/abs/1507.07493}{{\ttfamily 1507.07493}}].

\bibitem{Elder:2016yxm}
B.~Elder, J.~Khoury, P.~Haslinger, M.~Jaffe, H.~M\"uller and P.~Hamilton,
  \emph{{Chameleon Dark Energy and Atom Interferometry}},
  \href{https://doi.org/10.1103/PhysRevD.94.044051}{\emph{Phys. Rev. D}
  {\bfseries 94} (2016) 044051}
  [\href{https://arxiv.org/abs/1603.06587}{{\ttfamily 1603.06587}}].

\bibitem{Burrage:2016rkv}
C.~Burrage, A.~Kuribayashi-Coleman, J.~Stevenson and B.~Thrussell,
  \emph{{Constraining symmetron fields with atom interferometry}},
  \href{https://doi.org/10.1088/1475-7516/2016/12/041}{\emph{JCAP} {\bfseries
  12} (2016) 041} [\href{https://arxiv.org/abs/1609.09275}{{\ttfamily
  1609.09275}}].

\bibitem{Jaffe:2016fsh}
M.~Jaffe, P.~Haslinger, V.~Xu, P.~Hamilton, A.~Upadhye, B.~Elder et~al.,
  \emph{{Testing sub-gravitational forces on atoms from a miniature, in-vacuum
  source mass}}, \href{https://doi.org/10.1038/nphys4189}{\emph{Nature Phys.}
  {\bfseries 13} (2017) 938}
  [\href{https://arxiv.org/abs/1612.05171}{{\ttfamily 1612.05171}}].

\bibitem{Sabulsky:2018jma}
D.~O. Sabulsky, I.~Dutta, E.~A. Hinds, B.~Elder, C.~Burrage and E.~J. Copeland,
  \emph{{Experiment to detect dark energy forces using atom interferometry}},
  \href{https://doi.org/10.1103/PhysRevLett.123.061102}{\emph{Phys. Rev. Lett.}
  {\bfseries 123} (2019) 061102}
  [\href{https://arxiv.org/abs/1812.08244}{{\ttfamily 1812.08244}}].

\bibitem{Fischer:2023eww}
H.~Fischer, C.~K\"ading, H.~Lemmel, S.~Sponar and M.~Pitschmann, \emph{{Search
  for Dark Energy with Neutron Interferometry}},
  \href{https://doi.org/10.1093/ptep/ptae014}{\emph{PTEP} {\bfseries 2024}
  (2024) 023E02} [\href{https://arxiv.org/abs/2310.18109}{{\ttfamily
  2310.18109}}].

\bibitem{Mueller2024}
C.~D. Panda, M.~J. Trao, M.~Ceja, J.~Khoury, G.~M. Tino and H.~M\"uller,
  \emph{{Measuring gravitational attraction with a lattice atom
  interferometer}},
  \href{https://doi.org/10.1038/s41586-024-07561-3}{\emph{Nature} (2024) }
  [\href{https://arxiv.org/abs/2310.01344}{{\ttfamily 2310.01344}}].

\bibitem{Burrage2018}
C.~Burrage, C.~K\"ading, P.~Millington and J.~Min\'a\v{r}, \emph{{Open quantum
  dynamics induced by light scalar fields}},
  \href{https://doi.org/10.1103/PhysRevD.100.076003}{\emph{Phys. Rev. D}
  {\bfseries 100} (2019) 076003}
  [\href{https://arxiv.org/abs/1812.08760}{{\ttfamily 1812.08760}}].

\bibitem{Kading:2023mdk}
C.~K\"ading, M.~Pitschmann and C.~Voith, \emph{{Dilaton-induced open quantum
  dynamics}}, \href{https://doi.org/10.1140/epjc/s10052-023-11939-4}{\emph{Eur.
  Phys. J. C} {\bfseries 83} (2023) 767}
  [\href{https://arxiv.org/abs/2306.10896}{{\ttfamily 2306.10896}}].

\bibitem{Gasperini:2001pc}
M.~Gasperini, F.~Piazza and G.~Veneziano, \emph{{Quintessence as a runaway
  dilaton}}, \href{https://doi.org/10.1103/PhysRevD.65.023508}{\emph{Phys. Rev.
  D} {\bfseries 65} (2002) 023508}
  [\href{https://arxiv.org/abs/gr-qc/0108016}{{\ttfamily gr-qc/0108016}}].

\bibitem{Damour:2002nv}
T.~Damour, F.~Piazza and G.~Veneziano, \emph{{Violations of the equivalence
  principle in a dilaton runaway scenario}},
  \href{https://doi.org/10.1103/PhysRevD.66.046007}{\emph{Phys. Rev. D}
  {\bfseries 66} (2002) 046007}
  [\href{https://arxiv.org/abs/hep-th/0205111}{{\ttfamily hep-th/0205111}}].

\bibitem{Damour:2002mi}
T.~Damour, F.~Piazza and G.~Veneziano, \emph{{Runaway dilaton and equivalence
  principle violations}},
  \href{https://doi.org/10.1103/PhysRevLett.89.081601}{\emph{Phys. Rev. Lett.}
  {\bfseries 89} (2002) 081601}
  [\href{https://arxiv.org/abs/gr-qc/0204094}{{\ttfamily gr-qc/0204094}}].

\bibitem{Brax:2010gi}
P.~Brax, C.~van~de Bruck, A.-C. Davis and D.~Shaw, \emph{{The Dilaton and
  Modified Gravity}},
  \href{https://doi.org/10.1103/PhysRevD.82.063519}{\emph{Phys. Rev. D}
  {\bfseries 82} (2010) 063519}
  [\href{https://arxiv.org/abs/1005.3735}{{\ttfamily 1005.3735}}].

\bibitem{Brax:2011ja}
P.~Brax, C.~van~de Bruck, A.-C. Davis, B.~Li and D.~J. Shaw, \emph{{Nonlinear
  Structure Formation with the Environmentally Dependent Dilaton}},
  \href{https://doi.org/10.1103/PhysRevD.83.104026}{\emph{Phys. Rev. D}
  {\bfseries 83} (2011) 104026}
  [\href{https://arxiv.org/abs/1102.3692}{{\ttfamily 1102.3692}}].

\bibitem{Brax2022}
P.~Brax, H.~Fischer, C.~K\"ading and M.~Pitschmann, \emph{{The environment
  dependent dilaton in the laboratory and the solar system}},
  \href{https://doi.org/10.1140/epjc/s10052-022-10905-w}{\emph{Eur. Phys. J. C}
  {\bfseries 82} (2022) 934}
  [\href{https://arxiv.org/abs/2203.12512}{{\ttfamily 2203.12512}}].

\bibitem{Fischer:2023koa}
H.~Fischer, C.~K\"ading, R.~I.~P. Sedmik, H.~Abele, P.~Brax and M.~Pitschmann,
  \emph{{Search for environment-dependent dilatons}},
  \href{https://doi.org/10.1016/j.dark.2024.101419}{\emph{Phys. Dark Univ.}
  {\bfseries 43} (2024) 101419}
  [\href{https://arxiv.org/abs/2307.00243}{{\ttfamily 2307.00243}}].

\bibitem{Kading2022x}
C.~K\"ading and M.~Pitschmann, \emph{{New method for directly computing reduced
  density matrices}},
  \href{https://doi.org/10.1103/PhysRevD.107.016005}{\emph{Phys. Rev. D}
  {\bfseries 107} (2023) 016005}
  [\href{https://arxiv.org/abs/2204.08829}{{\ttfamily 2204.08829}}].

\bibitem{Fischer:2024gni}
H.~Fischer, C.~K\"ading and M.~Pitschmann, \emph{{Quantum and thermal pressures
  from light scalar fields}},
  \href{https://doi.org/10.1016/j.dark.2024.101756}{\emph{Phys. Dark Univ.}
  {\bfseries 47} (2025) 101756}
  [\href{https://arxiv.org/abs/2407.20658}{{\ttfamily 2407.20658}}].

\bibitem{Breuer2002}
H.-P. Breuer and F.~Petruccione, \emph{The Theory of Open Quantum Systems}.
  Oxford University Press, Oxford, 2002.

\bibitem{Bellac:2011kqa}
M.~L. Bellac, \emph{{Thermal Field Theory}}, Cambridge Monographs on
  Mathematical Physics. Cambridge University Press, 3, 2011,
  \href{https://doi.org/10.1017/CBO9780511721700}{10.1017/CBO9780511721700}.

\bibitem{Schwinger}
J.~S. Schwinger, \emph{{Brownian Motion of a Quantum Oscillator}},
  \href{https://doi.org/10.1063/1.1703727}{\emph{J. Math. Phys.} {\bfseries 2}
  (1961) 407}.

\bibitem{Keldysh}
L.~V. Keldysh, \emph{{Diagram technique for nonequilibrium processes}},
  {\emph{Zh. Eksp. Teor. Fiz.} {\bfseries 47} (1964) 1515}.

\bibitem{Feynman}
R.~P. Feynman and F.~L. Vernon, \emph{The theory of a general quantum system
  interacting with a linear dissipative system}, {\emph{Annals of physics}
  {\bfseries 24} (1963) 118}.

\bibitem{bcca-m22}
B.~Barrett, G.~Condon, L.~Chichet, L.~Antoni-Micollier, R.~Arguel, M.~Rabault
  et~al., \emph{Testing the universality of free fall using correlated
  39k–87rb atom interferometers}, {\emph{AVS Quantum Science} {\bfseries 4}
  (2022) 014401}.

\bibitem{eymkl15}
B.~Estey, C.~Yu, H.~Müller, P.-C. Kuan and S.-Y. Lan, \emph{{High-Resolution
  Atom Interferometers with Suppressed Diffraction Phases}}, {\emph{Phys. Rev.
  Lett.} {\bfseries 115} (2015) 083002}.

\bibitem{Lepoutre:2010zz}
S.~Lepoutre, H.~Jelassi, G.~Trenec, M.~Buchner and J.~Vigue, \emph{{Atom
  interferometry as a detector of rotation and gravitational waves: Comparison
  of various diffraction processes}},
  \href{https://doi.org/10.1007/s10714-010-1133-y}{\emph{Gen. Rel. Grav.}
  {\bfseries 43} (2011) 2011}.

\end{thebibliography}\endgroup
\bibliographystyle{JHEP}

\end{document}